\documentstyle[12pt,epsf]{article}
\begin{document} 
\begin{flushright} 
\begin{tabular}{l}
ULB--TH/96--5\\
hep-ph/9605241 \\
May 1995
\end{tabular} 
\end{flushright} 
\vskip1cm 
\begin{center}{\large{\bf Zeroes of the $\bf e^+ e^- \to \bar f f$ cross 
section 
\vskip0.5cm 
and search for new physics.}} 
\vskip1cm 
 
J.-M.Fr\`{e}re\footnote{postal address: Physique Theorique CP 225; U.L.B.
 Boulevard du Triomphe; B-1050 Bruxelles, Belgium;
email: frere@ulb.ac.be} \\
Universit\'e Libre de Bruxelles
\vskip1cm 
 
V.A.Novikov, M.I.Vysotsky\\ 
117259 ITEP, Moscow, Russia 
 
\end{center} 
 
\vskip5cm 
 
\noindent \underline{\bf Abstract}: 
\vskip1cm 
 
\noindent We suggest returning to a different presentation of the
$ e^+e^- \to 
\bar f f$ data off the $Z$ peak, with the hope of using zeroes of specific 
amplitudes to enhance the sensitivity to new physics. 
 
\newpage

\vskip0.5cm 
As well known the electron-positron annihilation into light fermions
$ e^+ e^- 
\to \bar f f$ is best described at tree level in terms of the helicity 
amplitudes, which do not interfere in the massless limit. 
 
\bigskip 
To the amplitudes $ A_{ij}^f (i, j = L, R)$ with $i$ referring to the 
polarisation of the initial electron, and $j$ to that of the final fermion 
correspond the cross sections given by $ \sigma_{ij}$, and in the 
tree level massless limit: 
 
\begin{eqnarray} 
\sigma = {1 \over 2}(\sigma_{LL} + \sigma_{RR} + \sigma_{LR} + \sigma_{RL})
\label{I}  \end{eqnarray} 
 
\bigskip 
\noindent with $ {1 \over 2}$ corresponding to the spin averaging
 for unpolarized initial beams.

\noindent 
Due to $ \gamma - Z$ interference, we expect each of these 
terms to vanish (up to the $Z$ width) for a specific value of the 
center of mass energy $ \sqrt s$. 
 
\bigskip 
For the known fermions, we observed that $ \sigma_{LL}$ and
$ \sigma_{RR}$ both 
vanish for $ \sqrt s < m_Z$, while the zeros in $ \sigma_{LR} + \sigma_{RL}$ 
are found at $ \sqrt s > m_Z$ with the exception of $ 
\sigma_{RL}^{d}$, for which the photon coupling is insufficient to achieve 
 a total destructive interference. Instead of the usual analysis in terms of 
forward-backward asymmetries $A_{FB}$, we advocate thus to turn to 
the above decomposition using appropriate filters. Separate studies of
$(\sigma_{LL} + \sigma_{RR})$ and $ (\sigma_{LR} + \sigma_{RL})$, or of their
ratios,  close to the $ \gamma Z$ interference minima allow to: 
 
\begin{itemize} 
\item exhibit in a self-explanatory way the $ \gamma - Z$ interference. 
\item test the position and height of the minima 
\item minimize the background for the observation of possible new scalar 
interactions (which don't interfere with the $Z$, and don't benefit 
in any case from the $Z$ peak enhancement) \item understand readily 
the different impact of extra gauge bosons $(Z^\prime)$, 
family-universal or otherwise . 
\end{itemize} 
 
\noindent The possibility also exists, at least at the theoretical level, to 
enhance the up- or down-quark content of the final state
(in particular $c$ and 
$b$ quarks). 
\bigskip 
 
\noindent Working with limited off-$Z$-peak statistics (which is {\it a 
fortiori} 
the case with the data obtained from hard initial state
radiation $ e^+ e^- \to 
f^+ f^- \gamma)$ we need to provide an efficient way to extract the helicity 
amplitudes. 
 
 \bigskip 
 
\noindent In the absence of polarization at LEP, the simplest is to use the
angular  dependence of the various terms: 
 
\begin{eqnarray} 
\sigma_{LL}, \sigma_{RR} &\sim& ( 1 + \cos \theta)^2 \nonumber \\ 
\sigma_{LR}, \sigma_{RL} &\sim&  (1 - \cos \theta)^2 \; . \label{II} 
\end{eqnarray} 
 
\bigskip 
 
\noindent If no other amplitudes are present (scalar interactions, higher loop 
corrections) and only these 2 terms contribute, a simple measurement of the 
integrated forward-backward asymmetry is in principle sufficient.
One could also 
consider projecting on the individual partial waves, (which even neglecting 
radiative corrections means 3 different polynomials). 

Keeping in mind a search for new physics, and the restricted
data available, our personal preference 
consists in \underline{projecting out} the non-vanishing $ \sigma_{ij}$ by a 
suitable filter $F$. The result of this projection is then constituted of 
contributions from the remaining amplitude at the minimum, radiative
corrections 
(which can be modelled) and any new physics. 
\vskip0.5cm 
 
\noindent For limited sets of data, this procedure seems considerably more 
stable than a full partial wave analysis. 
\noindent For instance, the $ LR + RL$ amplitude (respectively $RR + LL$) are 
eliminated by the filters 
 
\begin{eqnarray} 
\tilde \sigma_+ = \int_{-\cos \theta max} ^{\cos \theta max}
{d \sigma \over d 
\cos \theta} F_ + d \cos \theta  &=& \sigma_{LL} + \sigma _{RR}
\nonumber \\ 
&+& {\rm rad.corrections } +{\rm new\;  physics}, \nonumber \\ 
\tilde \sigma_-  = \int_{-\cos \theta max} ^{\cos \theta max}
{d \sigma \over d 
\cos \theta} F_ - d \cos \theta  &=& \sigma_{LR} + \sigma	_{RL} 
\nonumber \\ 
 &+& {\rm rad.corrections } +{\rm new\;  physics}.    \label{III} 
 \end{eqnarray} 
 (For $\cos \theta_{max} = 1$ we have $F_- = 1-2\cos\theta\;, \; F_+ 
 = 1 + 2\cos\theta$). 
 
\bigskip 
 
\noindent We will now turn successively to a discussion of the zeroes, the
effect  of new physics, and finally, comment on the inclusion of radiative
corrections,  which will be required if off-$Z$ peak data become abundant. 
 
\bigskip 
\noindent \underline {Discussion of the zeroes}

\noindent We have to consider the amplitudes for the annihilation between an 
electron of polarization $ i_1$, and an anti-(electron of polarization $i_2)$ 
into a fermion of pol.$j_1$ and an anti-(fermion of pol.$j_2)$
(the term of CP 
conjugate would be more appropriate, since a right-handed positron will be 
noted $(\bar e_L))$. 
 
\bigskip 
 
\noindent For instance: $(\bar e_L)  e_L \to (\bar \mu_R) \mu_R $
describes the annihilation of a right-handed positron with a 
left-handed electron into a right-handed muon and a left-handed antimuon and 
would be noted: $A_{LL; RR}$.  
 
\noindent Gauge vectors respect chirality and, as a result the corresponding
$A$  at tree level, will in the above notation always read $ A_{ij}$. 
 
\bigskip 
 
\noindent On the contrary, scalar (tensor) interactions have necessarily $i_1 
\neq i_2$ and $ j_1 \neq j_2$. As a direct consequence they cannot 
interfere with the above in the massless fermion limit. This has the
important 
consequence that a heavy scalar exchange signal would not benefit from 
interference with the  $Z$ close to the peak; on the contrary, 
the signal would  simply be drowned by the overwhelming $Z$ exchange. 
 
\bigskip 
 
\noindent 
Each of the helicity amplitudes is governed by the $ \gamma-Z$ 
interference, and we have in standard notations, but 
allowing for an extra gauge boson $Z^\prime$ for later use:
 
\begin{eqnarray} 
A_{ij}^f(q^2) &=& 
{-4\pi \bar{\alpha} Q_f \over q^2} + \sqrt{2}G_{\mu}m_Z^2 
 \;{g_i ^e g_j^f \over q^2 - (m_Z - {i \Gamma_z \over 2})^2} + 
{g^{\prime e}_i g_j^{\prime f} \over q^2 -(m_{Z^\prime} - 
{i\Gamma_{Z^\prime} \over 2})^2} \nonumber \\ &=& 
{-4\pi \bar{\alpha} Q_f \over 
q^2} \left[ 1 + R_{ij}^f (q^2) \right]  \label {VIII} 
\end{eqnarray} 
 
\bigskip 
 
\noindent 
where $ Q_f$ if the electric charge of the final state fermion, and $ 
g_i$ are the $L$ and $R$ coupling in the usual conventions. In the 
standard model:

\begin{eqnarray} 
g_L &=& 2 t_{3L} - 2 Q \sin^2 \theta\;\;, \nonumber \\ 
g_R &=& - 2 Q \sin^2 \theta \;\;, \nonumber \\ 
\bar{\alpha} &\equiv& \alpha(m^2_Z) \;\;, \nonumber \\ 
\sin^2 2\theta &=& {4\pi\bar{\alpha} \over \sqrt{2} G_{\mu} 
m^2_Z}\;\;. 
\label{IX} 
\end{eqnarray} 
 
\bigskip 
\noindent 
This parametrization uses the best known quantities $G_{\mu},~~ 
m_Z~~,~~ \bar{\alpha}$. Choice of the quantity 
$\bar{\alpha}$ takes 
into account the  main (electromagnetic) radiative corrections 
\cite{NOV}. 
 
Destructive interference will occur below the $ Z$  peak if 
 
\begin{eqnarray} 
(-  Q_f) /g^l_i  g_j^f > 0 \label {X} 
\end{eqnarray} 
 
\bigskip 
 
\noindent and above the $Z$ peak otherwise. It turns out in the
standard model 
that for all light fermions $(\mu, \tau, u, d ...)$ the destructive
interference 
(leading to near-zeroes) occurs for $\sqrt s < m_Z$ in the $ LL$ and $ RR$ 
channels (for leptons, this is obvious due to lepton universality), and for 
$\sqrt s > m_Z$ in the $LR + RL$ channels.(see table) \begin{center} 
\begin{tabular}{|c|c|c|c|c|} \hline 
particle type    &     LL       &    RR     &     LR   &   RL \\ 
\hline $\mu$    &    76.9    &    80.0    &    113.1   & 113.1 \\ 
\hline u    &    68.2    &    80.0    &    113.1 &   159.9 \\ 
\hline d      &    53.4    &    80.0    &    113.1  &    none \\ 
\hline \end{tabular} \end{center} \bigskip 
 
table 1: location of zeroes (in GeV, tree level in weak interactions); 
in calculations we use $m_Z=91.188$ GeV, $\sin^2\theta =0.2311$. 

The locations of zeroes are at prominent  values of energies: 
$m_W$ for $RR$ and $\sqrt{2}m_W$ for $LR$ (for any channel), 
$\sqrt{2}m_W$ 
for $RL$ for charged leptons.

\bigskip 
\noindent 
In practice the projected cross sections $ \tilde \sigma_+, 
\tilde\sigma_-$ do not reach a zero minimum. This for two reasons: 
 
\begin{itemize} 
\item[i)] we have (although this is technically beyond the tree level)
to take 
into account the width of the $Z$, which gives an imaginary part to the
sole $Z$ 
contribution; resulting in a residual cross-section $ \sim \Gamma_Z ^2$
at the 
minimum. 
\item[ii)] selecting definite amplitudes by projection does not allow to 
distinguish between $LL$ and $RR$; (eq.\ref{II}). Since the zeros in $ 
Re (A_{LL})$ and $ Re (A_{RR})$ occur at different points, $\tilde \sigma_+$ 
or  $\sigma_{LL} + \sigma_{RR}$ only sees the sum of these dips (still
a very noticeable effect). 

The location of the zero  for cross section $\sigma_{ij}$ is 
given by the following  equation: 
\begin{equation}
{m_Z^2 \over E_0^2} = 1 - {1 \over 4 s^2 c^2} {g_i^e g_j^f \over Q_f} 
\;\; , 
s^2 \equiv \sin^2 \theta\;, \;\; c^2 = \cos^2 \theta\;\;. 
\end{equation}

Their separation would require, in addition to filtering, the identification 
of one polarization (initial state at SLC or $\tau$ polarization at LEP). The 
same is true for quarks produced in the  $ LR + RL$ channel above the $Z$
peak. 
 
The only case where we can really approach a unique minimum is for leptons in 
the $LR + RL$ channel above the $Z$ peak. \footnote{ 
If the position of the minimum can be measured with high accuracy 
this will give  additional way to study electroweak radiative corrections.}
As we shall see however, the combination of 2 nearby minima still yields a
severe suppression of the signal!
\end{itemize}
 
\bigskip

\noindent \underline{Filters}

We have advocated above (eq.\ref{III}) the use of a simple linear filter
to project out the unwanted $(LL + RR)$ or $ (LR + RL)$ contributions.

\noindent As the reader will check easily, this choice of a simple linear 
filter is insufficient to eliminate for instance a hypothetical scalar
component: ($  A$ and $C_M$ are defined in eq.\ref{XIII} below) 

\begin{eqnarray} 
\int_{-\cos \theta_{max}}^{\cos \theta_{max}} S_S \quad F_\pm \quad d
\cos\theta  = A\cdot C_M \cdot \sigma_S \label {XI} 
\end{eqnarray} 
 
\bigskip 
\noindent This choice of filtering is however deliberate, for 3 reasons: 
 
\begin{itemize} 
\item[-] our purpose is to put new physics into evidence, and the way to 
achieve this is to filter out the unwanted (non-vanishing) component of the 
known 
amplitude; namely, below the $Z$ poles we want to filter out the $ LR + RL$ 
part, in order to examine the minimum of the $ RR + LL$ part, and put in 
evidence possible new contributions. ($\sigma_S$ is just one example, other 
terms are possible). 
\item[-] radiative corrections can be relatively easily handled in this
scheme 
(see below). 
\item[-] a more elaborate fit, projecting on all the possible amplitudes (e.g, 
a Legendre polynomial expansion to order $ \cos^2 \theta$) proves much less 
stable then the above simple prescription. 
\end{itemize} 
 
\bigskip 
 
\noindent 
The (differential) tree level cross sections corresponding to eq.\ref{VIII}
read: 
 
\begin{eqnarray} 
\label{Xbis} 
S_{ij} &=& N_C . {\alpha^2 \over 8 s} . 2 \pi \vert 1 + R_{ij}^f \vert^2 . 
Q_f^2 \nonumber \\ 
{d\sigma \over d \cos \theta} &=& {(S_{LR} + S_{RL}) \over 2} (1 -\cos
\theta)^2
+  {(S_{RR} + S_{LL}) \over 2} (1 + \cos \theta)^2 + S_S \nonumber \\ 
\sigma &=& {8 \over 3} {(S_{LR}+S_{RL}+S_{RR}+S_{LL}) \over 2} +2 S_{S} 
\end{eqnarray} 
 
\bigskip 
 
\noindent where we have added for the sake of further discussion a
hypothetical 
scalar exchange contribution. 
 
\noindent The filters introduced in eq.\ref{III} are defined so that: 
 
\begin{eqnarray} 
\label{XIbis} 
\int_{-\cos \theta_{max}} ^{\cos \theta_{max}} ( 1 \pm \cos \theta)^2 F_{\pm} 
 d \cos \theta &=& {8 \over 3} \nonumber \\ 
\int_{-\cos \theta_{max}} ^{\cos \theta_{max}} ( 1 \pm \cos \theta)^2
 F_{\mp} d \cos \theta &=& 0 
\end{eqnarray} 
 
\noindent 
which yields : 
 
\begin{eqnarray} 
F_{\pm}&=& A \left( 1 {\pm} B \cos \theta \right) \nonumber \\ 
A &=& {2 \over C_M (3 + C_M^{2})}  \nonumber \\ 
B &=&  {3 + C_M^{2} \over 2 C_M^{2}} \nonumber \\ 
C_M &=& \cos \theta _{max}  \label {XIII} 
\end{eqnarray}

\noindent In Figs 1 and 2, we plot the standard model expectation for 
 
\begin{equation} 
{\tilde\sigma_+ ^\mu \over \tilde \sigma_-^\mu} = { \int {d \sigma \over
d \cos 
\theta} . F_+ d \cos \theta \over \int {d \sigma \over d\cos
\theta} . F_- d \cos 
\theta}\label {XV} 
\end{equation} 
 
\bigskip 
 
\noindent for $ \sqrt s < m_Z$ and $ {\tilde \sigma_-^\mu \over 
\tilde \sigma_+^\mu}$ for $ \sqrt s > m_Z$. 
 
\noindent While, as previously discussed, the dip in
$ \sigma_{RR} + \sigma_{LL}$ 
results from the superposition  of 2 nearby minima,
$ \sigma_{LR+ RL}^\mu $ is 
easier to interpret and is proportional to $\Gamma_Z^2$. 
 \bigskip

\noindent We want to stress the relation between filtering and the 
forward-backward asymmetry $ A_{FB}$. According to the standard
definition, we get 
considering only the contributions in eq. \ref{Xbis} (more 
complicated terms are expected from radiative corrections)

\begin{equation}
A_{FB} = {3 \over 4} \; {\sigma_{LL} + \sigma_{RR}- \sigma_{LR} -
\sigma_{RL}\over \sigma_{LL} + \sigma_{RR}+ \sigma_{LR} +
\sigma_{RL} 
+ 2\ \sigma_S} 
\label{XVI}
\end{equation}

\bigskip
\noindent We note first that $ A_{FB}$ has only (and by design) little 
sensitivity
 to a possible scalar contribution, at the difference of
$ \tilde \sigma_+$ or 
$\tilde \sigma_-$ near to a zero.

\noindent Only in the case of pure single vector exchanges (SVE) (like the 
tree-level standard model) are the two presentations equivalent,
 with the simple 
relation:

\begin{equation}
{\tilde \sigma_+ \over \tilde \sigma_-} \vert _{SVE} = {1 + 4/3 \  A_{FB}
 \over 1 - 4/3 \  
 A_{FB}} \label{XVII}
\end{equation}

\bigskip 
 
\noindent \underline{Windows for new physics}. 

 One of the standard searches \cite{RR} is for 
 extra $Z^\prime$ neutral bosons. Such particles would simply add to 
 the existing amplitude, according to eq. \ref{VIII}. In the case of $ e^+ 
e^- \to \mu^+ \mu^-$ below the $Z$ peak, we want to concentrate on 
the $ LL + RR$ amplitudes. If the $Z^\prime$ couplings respect lepton 
universality $ (g_i^{\prime f} = g_i ^{\prime e})$ it is easy to see 
that they simply enhance the $Z$ effect, thereby bringing the minima 
to lower $ \sqrt s$; the resulting curve  dips below the standard one for  $ 
\sqrt s \ll m_Z$, but raises above it on the rising flank of the $Z$ 
peak. 
 
The opposite is observed if for instance $ (g_i^{\prime f} = - g_i^{\prime 
e})$, a situation expected in "horizontal" interactions. 
\bigskip 
 
We don't expect spectacular limits to arise for far off-peak data however, 
since extra gauge boson contributions interfere with the real part of the $ 
(\gamma, Z)$ amplitudes and are thus widely enforced when this is large: 
sitting on the minimum of the $ \gamma, Z$ curve minimises this effect, and 
gain in background reduction is unlikely to compensate for the loss in 
statistics. In passing however, we note that the analysis of ref. \cite{RR} 
would usefully be extended to non-universal couplings in order to allow both 
for constructive and destructive interference. 
\bigskip

Quite different is the situation for scalar boson exchanges. As stressed 
before, conservation of chirality prevents their interference wit
 gauge boson 
contributions in the massless fermion limit. The best place to look
for such a 
signal is at the minimum (zero) of the gauge contribution. Here, the filter 
technique has a clear advantage, as the forward-backward asymmetry is rather 
insensitive to scalars.
\bigskip

In quite a different domain, it is amusing to note that in the $ LR + RL$ 
projection of the $ \bar \mu \mu$ cross section, the minimum (up to radiative 
corrections) is proportional to $ \Gamma_Z^2 (s)$, thus allowing in
principle an 
independent determination of this quantity off the $Z$ peak.

Let us stress as well that this minimum occurs at $E =\sqrt{2} m_W$. So 
$m_W$ can be measured 
at LEP 1.5 with the accuracy close to that of $m_Z$  below 
$2W$ threshold (at least in principle).

The above considerations can obviously be translated (with greater
experimental 
difficulties) to $\tau$ leptons (where polarization identification
combined with 
projection would then allow a complete separation of the 4 
amplitudes!), and to the quark sector. 
 
We had some hope that the existence of different minima for up and down
quarks 
would allow for an easy separation of these contributions. Unfortunately, as 
seen from the table, such separation is only quantitative. For the sake of 
argument, we present in fig (3) the ratio $ \tilde \sigma_(d) /\tilde 
\sigma_(u)$ which shows the possibility for an enrichment for instance in
$b$ or 
$s$ versus $c$ quarks. Here also, the loss in statistics is such that this 
method could only be  of use as a last ditch effort to improve a sample's
purity. 
 
\bigskip 
 
\noindent \underline{Radiative corrections}: 
 
With increasing data, radiative corrections will be needed. 
 
\noindent Following standard usage, they are usefully divided in 3 groups: 
 
\begin{itemize} 
\item[-] electroweak corrections to the propagators and vertices;
these are most 
easily taken into account, and amount essentially to effective 
(momentum-dependent) coupling constants. 

The largest corrections of this type are of electromagnetic nature 
and have been taken into account already in eqs. \ref{VIII},\ref{IX}.

\item[-] initial and final state soft radiation; their treatment is usually 
experiment dependent; we can expect them to factorize so that the 
ratios $ \tilde \sigma_+ / \tilde \sigma_-$ would not be significally 
affected. 
 
\item[-] box diagrams are the most difficult new contributions, since they 
induce a much more complicated angular $ (\cos \theta)$ dependence. 
It seems both difficult and unnecessary to fit higher order polynomials in $ 
\cos \theta$. Instead, we suggest to simply apply the above filters to the
full theoretical cross sections using the existing 
codes \cite{B} and to compare the experimental curves to that result. 
\end{itemize} 

\noindent \underline{Conclusions} 
 
We have suggested a presentation of $ e^+ e^- \to \bar ff$ data, in terms of 
helicity instead of asymmetries, and suggested the use of a specific 
projection filter to achieve this goal. 
 
This presentation allows for a more transparent 
interpretation and understanding of the data, in particular where new physics 
searches is involved. 
 
We are grateful to  Catherine De Clercq for many enjoyable 
discussions. 
Our work was partially supported by the grant INTAS-94-2352; work of 
V.A.Novikov and M.I.Vysotsky was supported by the grant 
RFBR-96-02-18010 as well.

\begin{figure}
$$
\epsffile{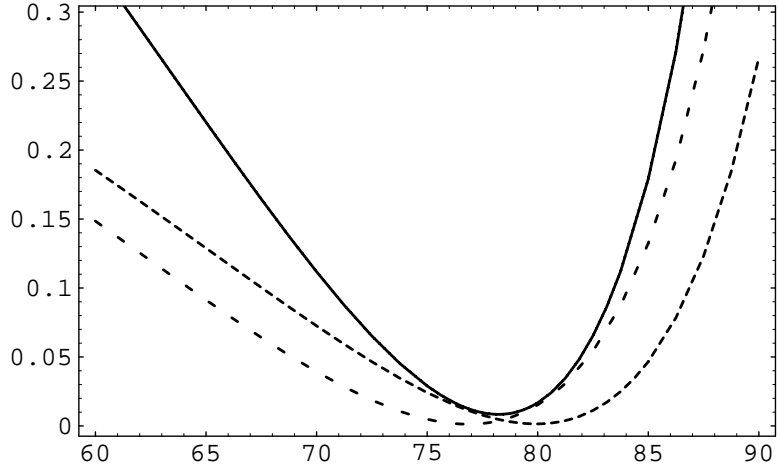}
$$

\vspace*{-1.0cm}
\caption[]{${{\sigma_{LL}}\over 
{(\sigma_{LR}+\sigma_{RL})}}$ (spaced  dashes),
${{\sigma_{RR}}\over{(\sigma_{LR}+\sigma_{RL})}}$ (close dashes),
${{\sigma_{LL}+\sigma_{RR}}\over{(\sigma_{LR}+\sigma_{RL})} }$
(solid line) 
as a function of center of mass 
energy (in GeV) 
for $e^+ e^- \to \mu^+ \mu^-$ below the Z peak.\label{fig1}}
\end{figure}
\begin{figure}
$$
\epsffile{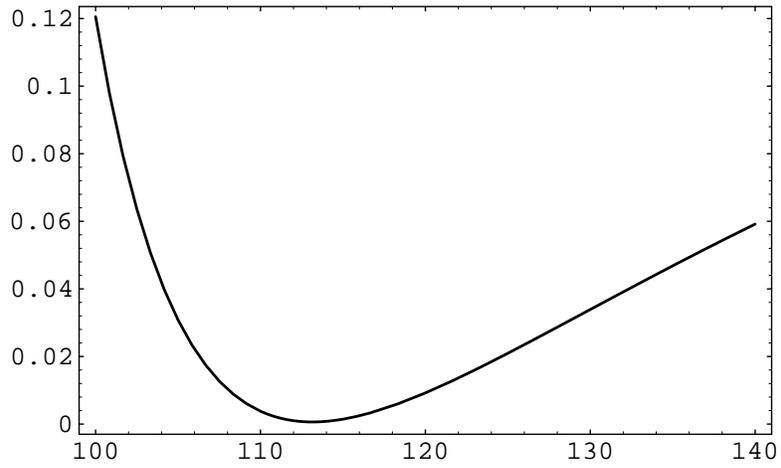}
$$

\vspace*{-1.0cm}
\caption[]{$(\sigma_{LR}+\sigma_{RL})\over 
{(\sigma_{LL}+\sigma_{RR})}$ as a function of center of mass
 energy (in GeV)  for $e^+ e^- \to \mu^+ \mu^-$ above the
Z peak.\label{fig2}}
\end{figure}

\begin{figure}
$$
\epsffile{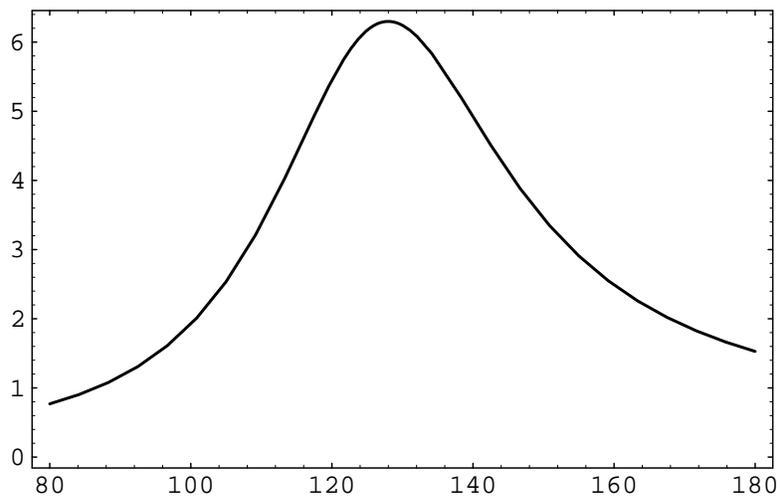}
$$

\vspace*{-1.0cm}
\caption[]{${\sigma^{d}_{LR+RL}}\over {\sigma^{u}_{LR+RL}}
$ as a function of center of mass 
energy (in GeV); the enhancement of d quarks production with respect 
to u quarks
is clearly seen in this channel. \label{fig3}}
\end{figure}

\end{document}